\documentclass[aps,prl,twocolumn,superscriptaddress,showpacs]{revtex4}
\usepackage{epsfig}

\begin{document}


\title{Stable Operation of a 300-m Laser Interferometer 
with Sufficient Sensitivity to
Detect Gravitational-Wave Events within our Galaxy}

\author{Masaki Ando}
\email[e-mail: ]{ando@granite.phys.s.u-tokyo.ac.jp}
\affiliation{Department of Physics, University of Tokyo, 
     7-3-1 Hongo, Bunkyo-ku, Tokyo 113-0033, Japan}

\author{Koji      Arai      }
\affiliation{National Astronomical Observatory of Japan,
  Mitaka, Tokyo 181-8588, Japan}

\author{Ryutaro   Takahashi }
\affiliation{National Astronomical Observatory of Japan,
  Mitaka, Tokyo 181-8588, Japan}

\author{Gerhard Heinzel     }
\affiliation{National Astronomical Observatory of Japan,
  Mitaka, Tokyo 181-8588, Japan}

\author{Seiji     Kawamura }
\affiliation{National Astronomical Observatory of Japan,
  Mitaka, Tokyo 181-8588, Japan}

\author{Daisuke   Tatsumi   }
\affiliation{National Astronomical Observatory of Japan,
  Mitaka, Tokyo 181-8588, Japan}

\author{Nobuyuki  Kanda     }
\affiliation{Department of Physics, Miyagi University of Education,
   Aoba Aramaki, Sendai 980-0845, Japan}

\author{Hideyuki  Tagoshi  }
\affiliation{Department of Earth and Space Science, Osaka University,
   Toyonaka, Osaka 560-0043, Japan}

\author{Akito     Araya     }
\affiliation{Earthquake Research Institute, University of Tokyo,
Bunkyo-ku, Tokyo 113-0032, Japan}

\author{Hideki    Asada    }
\affiliation{Faculty of Science and Technology, Hirosaki University,
Hirosaki, Aomori 036-8561, Japan}

\author{Youich Aso}
\affiliation{Department of Physics, University of Tokyo, 
     7-3-1 Hongo, Bunkyo-ku, Tokyo 113-0033, Japan}

\author{Mark A. Barton}
\affiliation{Institute for Cosmic Ray Research, University of Tokyo,
   Kashiwa, Chiba 277-8582, Japan}

\author{Masa-Katsu Fujimoto }
\affiliation{National Astronomical Observatory of Japan,
  Mitaka, Tokyo 181-8588, Japan}

\author{Mitsuhiro Fukushima }
\affiliation{National Astronomical Observatory of Japan,
  Mitaka, Tokyo 181-8588, Japan}

\author{Toshifumi Futamase  }
\affiliation{Astronomical Institute, Tohoku University,
Sendai, Miyagi 980-8578, Japan}

\author{Kazuhiro    Hayama }
\affiliation{Department of Astronomy, University of Tokyo,
     Bunkyo-ku, Tokyo 113-0033, Japan}

\author{Gen'ichi  Horikoshi }
\thanks{deceased}
\affiliation{High Energy Accelerator Research Organization,
   Tsukuba, Ibaragi 305-0801, Japan}

\author{Hideki    Ishizuka  }
\affiliation{Institute for Cosmic Ray Research, University of Tokyo,
   Kashiwa, Chiba 277-8582, Japan}

\author{Norihiko  Kamikubota}
\affiliation{High Energy Accelerator Research Organization,
   Tsukuba, Ibaragi 305-0801, Japan}

\author{Keita     Kawabe    }
\affiliation{Department of Physics, University of Tokyo, 
     7-3-1 Hongo, Bunkyo-ku, Tokyo 113-0033, Japan}

\author{Nobuki    Kawashima }
\affiliation{Department of Physics, Kinki University,
Higashi-Osaka, Osaka 577-8502, Japan}

\author{Yoshinori   Kobayashi}
\affiliation{Department of Physics, University of Tokyo, 
     7-3-1 Hongo, Bunkyo-ku, Tokyo 113-0033, Japan}

\author{Yasufumi  Kojima    }
\affiliation{Department of Physics, Hiroshima University,
Higashi-Hiroshima, Hiroshima 739-8526, Japan}

\author{Kazuhiro  Kondo  }
\affiliation{Institute for Cosmic Ray Research, University of Tokyo,
   Kashiwa, Chiba 277-8582, Japan}

\author{Yoshihide Kozai     }
\affiliation{National Astronomical Observatory of Japan,
  Mitaka, Tokyo 181-8588, Japan}

\author{Kazuaki   Kuroda    }
\affiliation{Institute for Cosmic Ray Research, University of Tokyo,
   Kashiwa, Chiba 277-8582, Japan}

\author{Namio     Matsuda   }
\affiliation{Department of Materials Science and Engineering, Tokyo Denki
University, Chiyoda-ku, Tokyo 101-8457, Japan}

\author{Norikatsu Mio       }
\affiliation{Department of Advanced Materials Science, University of Tokyo,
   Bunkyo-ku, Tokyo 113-0033, Japan}

\author{Kazuyuki  Miura     }
\affiliation{Department of Physics, Miyagi University of Education,
   Aoba Aramaki, Sendai 980-0845, Japan}

\author{Osamu     Miyakawa  }
\affiliation{Institute for Cosmic Ray Research, University of Tokyo,
   Kashiwa, Chiba 277-8582, Japan}

\author{Shoken M. Miyama    }
\affiliation{National Astronomical Observatory of Japan,
  Mitaka, Tokyo 181-8588, Japan}

\author{Shinji    Miyoki    }
\affiliation{Institute for Cosmic Ray Research, University of Tokyo,
   Kashiwa, Chiba 277-8582, Japan}

\author{Shigenori Moriwaki }
\affiliation{Department of Advanced Materials Science, University of Tokyo,
   Bunkyo-ku, Tokyo 113-0033, Japan}

\author{Mitsuru   Musha     }
\affiliation{Institute for Laser Science, University of Electro-Communications, 
   Chofugaoka, Chofu, Tokyo 182-8585, Japan}

\author{Shigeo    Nagano    }
\affiliation{Max-Planck-Institut f\"ur Quantenoptik, 
    Callinstrasse 38, D-30167 Hannover, Germany}

\author{Ken'ichi  Nakagawa }
\affiliation{Institute for Laser Science, University of Electro-Communications, 
   Chofugaoka, Chofu, Tokyo 182-8585, Japan}

\author{Takashi   Nakamura  }
\affiliation{Yukawa Institute for Theoretical Physics, Kyoto University,
    Kyoto 606-8502, Japan}

\author{Ken-ichi  Nakao     }
\affiliation{Department of Physics, Osaka City University,
   Sumiyoshi-ku, Osaka, Osaka 558-8585, Japan}

\author{Kenji     Numata    }
\affiliation{Department of Physics, University of Tokyo, 
     7-3-1 Hongo, Bunkyo-ku, Tokyo 113-0033, Japan}

\author{Yujiro    Ogawa     }
\affiliation{High Energy Accelerator Research Organization,
   Tsukuba, Ibaragi 305-0801, Japan}

\author{Masatake  Ohashi    }
\affiliation{Institute for Cosmic Ray Research, University of Tokyo,
   Kashiwa, Chiba 277-8582, Japan}

\author{Naoko     Ohishi    }
\affiliation{National Astronomical Observatory of Japan,
  Mitaka, Tokyo 181-8588, Japan}

\author{Satoshi   Okutomi    }
\affiliation{Institute for Cosmic Ray Research, University of Tokyo,
   Kashiwa, Chiba 277-8582, Japan}

\author{Ken-ichi  Oohara    }
\affiliation{Department of Physics, Niigata University,
     Niigata, Niigata 950-2102, Japan}

\author{Shigemi   Otsuka   }
\affiliation{Department of Physics, University of Tokyo, 
     7-3-1 Hongo, Bunkyo-ku, Tokyo 113-0033, Japan}

\author{Yoshio    Saito     }
\affiliation{High Energy Accelerator Research Organization,
   Tsukuba, Ibaragi 305-0801, Japan}

\author{Misao     Sasaki   }
\affiliation{Department of Earth and Space Science, Osaka University,
   Toyonaka, Osaka 560-0043, Japan}

\author{Shuichi   Sato     }
\affiliation{Institute for Cosmic Ray Research, University of Tokyo,
   Kashiwa, Chiba 277-8582, Japan}

\author{Atsushi   Sekiya    }
\affiliation{Department of Physics, University of Tokyo, 
     7-3-1 Hongo, Bunkyo-ku, Tokyo 113-0033, Japan}

\author{Masaru    Shibata   }
\affiliation{Department of Earth and Space Science, Osaka University,
   Toyonaka, Osaka 560-0043, Japan}

\author{Kentaro   Somiya    }
\affiliation{Department of Advanced Materials Science, University of Tokyo,
   Bunkyo-ku, Tokyo 113-0033, Japan}

\author{Toshikazu Suzuki    }
\affiliation{High Energy Accelerator Research Organization,
   Tsukuba, Ibaragi 305-0801, Japan}

\author{Akiteru   Takamori  }
\affiliation{Department of Physics, University of Tokyo, 
     7-3-1 Hongo, Bunkyo-ku, Tokyo 113-0033, Japan}

\author{Takahiro  Tanaka    }
\affiliation{Yukawa Institute for Theoretical Physics, Kyoto University,
    Kyoto 606-8502, Japan}

\author{Shinsuke  Taniguchi }
\affiliation{Department of Physics, University of Tokyo, 
     7-3-1 Hongo, Bunkyo-ku, Tokyo 113-0033, Japan}

\author{Souichi   Telada    }
\affiliation{National Research Laboratory of Metrology,
   Tsukuba, Ibaragi 305-8563, Japan}

\author{Kuniharu  Tochikubo }
\affiliation{Department of Physics, University of Tokyo, 
     7-3-1 Hongo, Bunkyo-ku, Tokyo 113-0033, Japan}

\author{Takayuki  Tomaru    }
\affiliation{Institute for Cosmic Ray Research, University of Tokyo,
   Kashiwa, Chiba 277-8582, Japan}

\author{Kimio     Tsubono   }
\affiliation{Department of Physics, University of Tokyo, 
     7-3-1 Hongo, Bunkyo-ku, Tokyo 113-0033, Japan}

\author{Nobuhiro  Tsuda     }
\affiliation{Precision Engineering Division, Tokai University, 
   Hiratsuka, Kanagawa 259-1292, Japan}

\author{Takashi   Uchiyama  }
\affiliation{High Energy Accelerator Research Organization,
   Tsukuba, Ibaragi 305-0801, Japan}

\author{Akitoshi  Ueda      }
\affiliation{National Astronomical Observatory of Japan,
  Mitaka, Tokyo 181-8588, Japan}

\author{Ken-ichi  Ueda      }
\affiliation{Institute for Laser Science, University of Electro-Communications, 
   Chofugaoka, Chofu, Tokyo 182-8585, Japan}

\author{Koichi    Waseda    }
\affiliation{National Astronomical Observatory of Japan,
  Mitaka, Tokyo 181-8588, Japan}

\author{Yuko      Watanabe  }
\affiliation{Department of Physics, Miyagi University of Education,
   Aoba Aramaki, Sendai 980-0845, Japan}

\author{Hiromi    Yakura   }
\affiliation{Department of Physics, Miyagi University of Education,
   Aoba Aramaki, Sendai 980-0845, Japan}

\author{Kazuhiro  Yamamoto  }
\affiliation{Department of Physics, University of Tokyo, 
     7-3-1 Hongo, Bunkyo-ku, Tokyo 113-0033, Japan}

\author{Toshitaka Yamazaki }
\affiliation{National Astronomical Observatory of Japan,
  Mitaka, Tokyo 181-8588, Japan}

\collaboration{the TAMA collaboration}

\date{\today}

\begin{abstract}
TAMA300, an interferometric gravitational-wave detector with 300-m 
baseline length, has been developed and operated with sufficient 
sensitivity to detect gravitational-wave events within our galaxy and 
sufficient stability for observations;
the interferometer was operated for over 10 hours stably and continuously.
With a strain-equivalent noise level of $ h\sim 5 \times 10^{-21}\,
/\sqrt{\rm Hz}$,
a signal-to-noise ratio (SNR) of 30 is expected for gravitational waves 
generated by a coalescence of 1.4\,$M_\odot$-1.4\,$M_\odot$ binary
neutron stars at 10\,kpc distance.
We evaluated the stability of the detector sensitivity
with a 2-week data-taking run, collecting 160 hours of data 
to be analyzed in the search for gravitational waves.
\end{abstract}


\maketitle


{\it Introduction.} ---
The direct observation of gravitational waves (GW) is expected to reveal 
new aspects of the universe \cite{bun-Thorne}.
Since GWs are emitted by the coherent bulk motion of matter,
and are hardly absorbed or scattered, they carry 
different information from that of electromagnetic waves.
However, no GW has yet been detected directly
because of its weakness.
In order to create a new field of GW astronomy,
several groups around the world are developing laser interferometric
GW detectors.
Compared with resonant-type GW detectors \cite{bun-reso},
interferometric detectors have an advantage in that they can observe 
the waveform of a GW, which would contain 
astronomical information.

Interferometric GW detectors are based on a Michelson interferometer.
The quadrupole nature of a GW causes differential changes in the arm
lengths of the Michelson interferometer, which are detected as changes 
in the interference fringe.
Interferometric detectors have been investigated with
many table-top \cite{bun-tabletop} and prototype \cite{bun-prototype} 
experiments to evaluate the principle of GW detection 
and their potential sensitivity to GWs.
With the knowledge obtained from these experimental interferometers, 
several GW detectors with baseline lengths of 300\,m to 4\,km
are under construction:
LIGO \cite{bun-LIGO} in U.S.A., 
VIRGO \cite{bun-VIRGO} and GEO \cite{bun-GEO} in Europe,
and TAMA \cite{mando-TAMA} in Japan.
In these detectors, both high sensitivity and high stability are 
required because the GW signals are expected to be extremely small 
and rare.

TAMA is a Japanese project to construct and operate
an interferometric GW detector with a 300-m baseline length
at the Mitaka campus of the National Astronomical Observatory in Tokyo
(${\rm 35^{\circ} 40' N,\ 139^{\circ} 32' E}$).
In this article, we report on an important achievement in  
interferometric detectors:
the TAMA detector was operated with sufficient sensitivity
and stability to observe GW events at the center of our galaxy.
The interferometer was operated stably and continuously over several
hours in typical cases, and over 10~hours in the best cases.
The noise-equivalent sensitivity was 
$ h\sim 5\times 10^{-21}\ /\sqrt{\rm Hz}$ at the floor level
(700\,Hz to 1.5\,kHz).
With this stability of the sensitivity, TAMA has the ability to detect 
GW events throughtout much of our galaxy: chirp signals from the coalescence 
of binary neutron stars or binary MACHO black holes \cite{mando-MACHO}, 
and burst signals from supernova explosions.

\begin{figure}[t]
  \begin{center}
  \epsfig{file=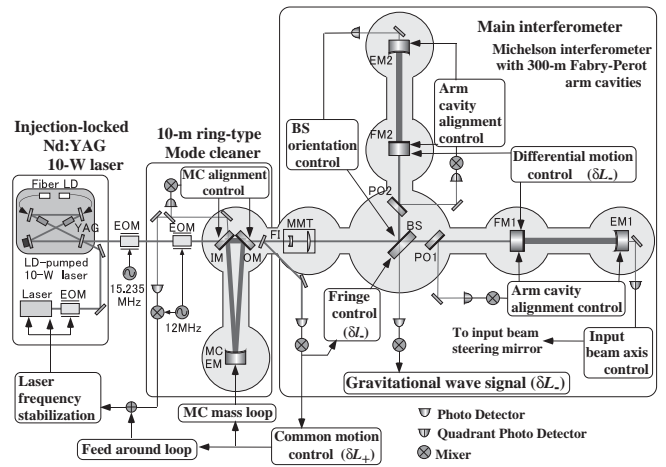, width=8.6cm}
  \caption{Optical and control design of the TAMA300 interferometer.
      TAMA300 is a Fabry-Perot-Michelson
      interferometer with a baseline length of 300\,m.
      A triangular ring cavity is inserted between the main interferometer
      and the laser source as a mode cleaner.
      The control system is designed to realize high sensitivity and
       stability of the detector at the same time.}
   \label{mando-diarec}
   \end{center}
\end{figure}

{\it Detector configuration.} ---
In the interferometer, called TAMA300, the arms of the Michelson
interferometer
are replaced by 300\,m Fabry-Perot arm cavities to enhance 
the sensitivity to GWs (Fig.\,\ref{mando-diarec}).
The arm cavities have a finesse of around 500, and 
a cutoff frequency of about 500\,Hz;
the light is stored in the cavities for about 0.3\,msec.
Since a high-power and stable laser is required as a light source, 
we use an LD-pumped Nd:YAG laser with an output power of 10\,W 
\cite{mando-10Wlaser1}.
In addition, a mode cleaner is inserted between the laser source and
the main interferometer to reject higher-mode beams and to stabilize the 
laser frequency.
The mode cleaner of TAMA300 is an independently-suspended
triangular ring cavity with a length of 9.75\,m \cite{mando-10mMC}.
Electro-optic modulators (EOM) for phase modulation 
(for mode cleaner and the main interferometer control) are placed
in front of the mode cleaner.
Thus, the wave-front distortion by the EOM
are rejected by the mode cleaner
before entering into the main interferometer. 

The mirrors of the main interferometer are made of fused silica.
Each mirror has a diameter of 100\,mm, and a thickness of 60\,mm.
The mirrors are coated 
by an IBS (ion-beam sputtering) machine to realize low-optical-loss
surfaces \cite{mando-mirrorloss2}.

The mirrors of the main interferometer and the mode cleaner are isolated 
from seismic motion by over 165\,dB (at 150\,Hz)
with three-stage stacks \cite{mando-stack} and double-pendulum 
suspension systems \cite{mando-sus}.
The suspension points are fixed to motorized stages, which are used for
an initial adjustment of the mirror orientations.
The fine position and orientation of each mirror is controlled with
coil-magnet actuators; small permanent magnets are attached to the mirror.

The interferometer is housed in a vacuum system
comprising eight chambers connected with beam tubes with a diameter
of 400\,mm.
With surface processing, called ECB (Electro-Chemical Buffing), 
a vacuum pressure of less than $ 10^{-6}$\,Pa
is achieved without baking \cite{mando-vac2}.

The control system is designed to realize high sensitivity and stability
at the same time.
It consists of three parts:
a length control system to keep the interferometer at its operational point,
an alignment control system to realize short-term ($\sim$\,1 minutes) stability
and high sensitivity,
and a beam-axis drift control system for long-term
($\sim$ a few hours) stable operation.
A frontal modulation scheme \cite{mando-FM} is used for the length control;
15.235\,MHz phase modulation is used for signal extraction.
The differential motion signal of the arm cavities ($\delta L_-$) is fed back to
the front mirrors with a bandwidth of 1\,kHz.
The common motion signal ($\delta L_+$) is fed back to the 
mode cleaner and the laser source to stabilize the laser frequency.
The motion in the Michelson interferometer part ($\delta l_-$)
is fed back to the beam splitter. 

An alignment control system is necessary for stable and sensitive operation
because angular fluctuations (about several $\mu$rad) of the suspended mirrors 
excited by seismic motion make the interferometer unstable.
The control signals are extracted by a wave-front-sensing scheme
\cite{mando-wfs}, and fed back to each mirror; 
the angular motions are suppressed by over 40\,dB to 
$ 10^{-8}$\,rad in root-mean-square.

Low-frequency drift control of the laser beam axis plays an important role
in maintaining long-term operation.
The beam axes are controlled with 300\,m optical levers;
the beam positions of the light transmitted through the arm cavities are 
monitored with quadrant photo detectors, and are fed back to the input 
steering mirror and the beam splitter of the main interferometer.

The data-acquisition system comprises a high-frequency part for
the main signals and a low-frequency part for detector diagnostics. 
The main output signals of the interferometer are recorded with
high-frequency A/D converters ($2\times 10^4$ samples/sec, 16\,bit)
after passing through whitening filters and 5\,kHz anti-aliasing 
low-pass filters.
Seven channel signals are recorded together with a timing signal, 
which provides a GPS-derived
coordinated universal time (UTC) within an accuracy of 1\,$\mu$sec.
Along with the high-frequency system, 88 channels of monitoring signals 
are collected with a low-frequency data-acquisition system
for interferometer diagnosis.

\begin{figure}[t]
\begin{center}
  \epsfig{file=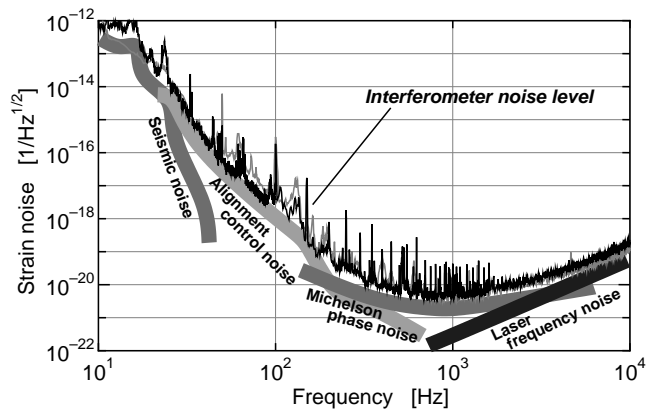, width=8.6cm}
  \caption{Noise level of the TAMA300 interferometer (black curve) 
    and the total contribution of identified noise sources 
     (gray curve).
  The floor level is $ 5\times 10^{-21}\ {\rm /\sqrt{Hz}}$ in strain.
  The thick curves represent the contribution of the noise sources.}
   \label{mando-sens2}
\end{center}
\end{figure}

{\it Detector noise level.} ---
Figure\,\ref{mando-sens2} shows the typical noise level of TAMA300
(black curve).
The displacement noise level of the interferometer is
$1.5\times 10^{-18}\ {\rm m/\sqrt{Hz}}$,
which corresponds to $ 5\times 10^{-21}\ {\rm /\sqrt{Hz}}$ in strain.
Almost all of the noise sources which limit the interferometer noise level
have been identified.
The gray curve in Fig.\,\ref{mando-sens2} represents the 
total contribution of the identified noise sources:
seismic motion ($\sim$\,30\,Hz), alignment-control noise
(30\,Hz$\,\sim\,$300\,Hz), 
Michelson phase-detection noise (300\,Hz$\,\sim$\,3\,kHz),
and the laser frequency noise (3\,kHz\,$\sim$).
The seismic noise and the laser frequency noise are estimated to satisfy
the design requirements in the observation band (around 300\,Hz).

Though the alignment control system is indispensable for
the stable operation of the interferometer,
this system can introduce excess noise to the interferometer
\cite{mando-ali}.
The noise in the alignment control error signal causes displacement noise
by coupling with mis-centering of the beam on a mirror,
and efficiency asymmetries of the coil-magnet actuators on a mirror.
In order to reduce this noise,
the actuator balances are adjusted so that the rotational center 
is at the beam spot on a mirror.
In addition, the beam position on each mirror is controlled to be 
still by the beam-axis control system.

The noise floor level is limited by the phase-detection noise of the 
Michelson part of the interferometer.
Phase changes in the light caused by GW are amplified
in the arm cavities, and detected by the Michelson interferometer.
In order to realize the designed detector sensitivity,
the noise level of the Michelson part of the interferometer should
be limited only by the shot noise in the observation band.
However, in TAMA, as well as by the shot noise, 
it is currently limited by the scattered light 
noise caused by the anti-reflection coating of the mode-matching telescope 
(MMT in Fig.\,\ref{mando-diarec})
between the main interferometer and the mode cleaner.
To realize the designed detector sensitivity which is purely limited by
the shot noise, the scattering noise should be removed.

\begin{figure}[t]
\begin{center}
  \epsfig{file=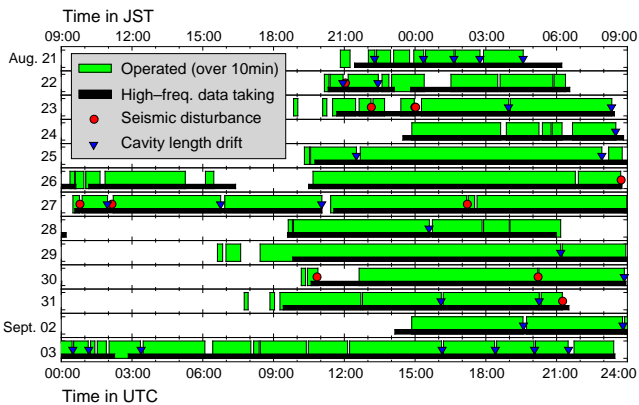, width=8.6cm}
  \caption{Operation status of the TAMA interferometer
     during a data-taking run performed from August 21 to September 3 in 
     the year 2000.
     The interferometer was operated stably for over 160 hours, 
     94.8\% of the total data-taking-run time.}
   \label{mando-lock}
\end{center}
\end{figure}

{\it Stability of operation.} ---
With the total system described above,
we performed a 2-week observation run
from August 21 to September 3, 2000.
Figure~\ref{mando-lock} shows the operational state during the observation;
the gray and black boxes represent the time when the interferometer was 
operated and when the data were taken, respectively.
The time is shown both in UTC and in Japan standard time (JST).
The interferometer was operated for over 160 hours, 
94.8\% of the total data-taking-run time.
(Periods of continuous lock shorter than 10\,minutes are not included.)
We operated the interferometer mainly during the night 
for the efficient collection of high-quality data.
During the observation, the noise level was degraded slightly 
from the Fig.\,\ref{mando-sens2} level, because of electronic noises 
by many cables connected to the interferometer for data-taking.
The longest continuous locking time was over 12 hours
(several hours in typical cases);
the main cause of loss of lock of the interferometer was
large seismic disturbances, including earthquakes (closed circles), 
and rather large drifts of $\delta L_+$ (triangle marks).
From a coincidence analysis of signals  from seismometers placed 
at the center and end rooms, we found that the interferometer 
was knocked out of lock by accelerations of 12\,mgal 
(1\,$\sim$\,10\,Hz frequency range).
The interferometer was also knocked out of lock 
by large drift of $\delta L_+ $;
a drift of over 120\,$\mu$m could cause saturation in the feedback loop.
The other causes of loss of lock are thought to be local seismic 
disturbances caused by human activities, spikes due to
instability of the laser source, and so on.

The interferometer noise level was stable thanks to the alignment and
drift control systems;
the drift of the noise level averaged for 1 minute was kept 
within a few dB, typically.
During the 2-week data-taking run,
the noise floor-level drift was kept within 3\,dB for
about 90\% of the total operation time.
In addition, in typical cases, the noise level was easily recovered 
without any manual adjustment after the unlock and relock 
of the interferometer with these automatic control systems.
The noise level of the interferometer was calibrated continuously 
using a sinusoidal calibration signal at 625\,Hz;
from the amplitude and phase of this peak signal, we 
estimated the optical gain and cut-off frequency of the cavity.
The Gaussianity of the noise level was evaluated  
every 30 seconds.
We observed about 10 non-Gaussian (confidence level of 99\%) events 
per hour in this observation run.
The non-Gaussian noise will be partly removed by veto analysis using other
channels, such as seismic motion, laser intensity noise, contrast 
fluctuation, and a $\delta L_+$ signal.

\begin{figure}[t]
\begin{center}
  \epsfig{file=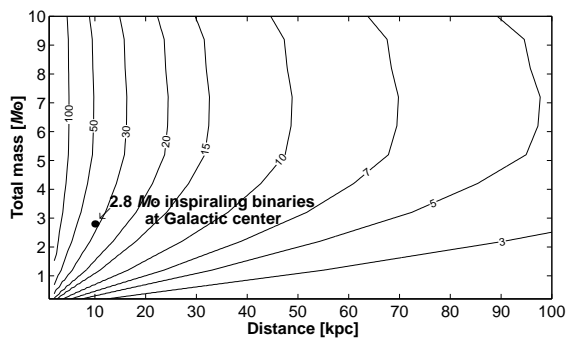, width=7.5cm}
  \caption{Contour plot of expected SNR
           for GW from inspiraling compact binaries with equal mass.
           Optimally-polarized GWs from an optimal direction for
           the detector are assumed.
           TAMA has the sensitivity to detect 1.4\,$M_\odot$-1.4\,$M_\odot$
           binary coalescence at Galactic center with SNR of 30.}
   \label{mando-sens}
\end{center}
\end{figure}

In order to check the GW-detection ability of TAMA,
we calculated the expected SNR for GW from
inspiraling binaries with the interferometer noise spectrum
and calculated chirp signals (Fig.\,\ref{mando-sens})
\cite{mando-Tago}.
Here, we assumed optimally-polarized GWs from an optimal direction for
the detector.
TAMA would detect GW events at the Galactic center
with sufficient SNR;
the SNR is about 30 in the case of chirp signals from a coalescence of
1.4\,$M_\odot$-1.4\,$M_\odot$ binary neutron stars at 10\,kpc distance.
With the burst signals from supernova explosions, 
TAMA would detect GWs with a strain 
amplitude of $h_{\rm rms} \sim 1 \times 10^{-18}$ 
(which corresponds to a mass energy of $\sim 0.01 M_\odot$,
again at the distance to the Galactic center)
with a SNR of about 10 at the frequency band from 700\,Hz to 1\,kHz.
However, as well as a veto analysis with the other recorded 
channels, a coincidence analysis with other GW detectors or other 
astronomical channels will be required for the detection to reject
non-Gaussian noise background.

{\it Conclusion.} ---
The TAMA300 interferometer has been operated stably for 
over 10 hours without loss of lock, with a noise-equivalent sensitivity of
$h\sim 5 \times 10^{-21}\ {\rm /\sqrt{Hz}}$ at the floor level.
With this sensitivity and stability, TAMA has the ability to
detect GW events within our galaxy,
though such events are expected to be very rare.
In order to increase the detection probability
for GW events farther away from our galaxy,
we are improving the detector sensitivity and stability
further.
Almost all of the noise sources which limit the detector sensitivity have 
been identified.
The noise level will be improved with new alignment control filters
and a reflective mode-matching telescope.
The stability of the operation will be improved further
with installation of an active isolation system and 
replacement of a suspension system by one with effective damping.

The achieved performance of the TAMA detector is a significant 
milestone in the quest for direct detection of GW, and for the 
establishment of GW astronomy with interferometric detectors.


The TAMA project is supported by a Grant-in-Aid for Creative Basic Research 
from the Ministry of Education.

\end{document}